\documentstyle[12pt,epsf]{article}

\begin{document}
 
\begin{titlepage}

\title{
\hfill
\parbox{5cm}{\normalsize 
\normalsize UT-803\\
}\\
\vspace{2ex}\large
Critical Exponents and Critical Amplitude Ratio of The Scalar
Model from Finite-temperature Field Theory 
\vspace{2ex}}
\author{\large
Kenzo Ogure\thanks{e-mail address:
 {\tt ogure@icrhp3.icrr.u-tokyo.ac.jp}}\\
 {\it Institute for Cosmic Ray Research,
   University of Tokyo, Midori-cho,}\\
   {\it   Tanashi, Tokyo 188, Japan}\\
and\\
Joe Sato\thanks{e-mail address:
 {\tt joe@hep-th.phys.s.u-tokyo.ac.jp}}\\
 {\it Department of Physics, School of Science,
   University of Tokyo,}\\
   {\it   Tokyo 113, Japan}}
\date{\today}

\maketitle

\begin{abstract}
    The critical exponents and the critical amplitude ratio of the
    scalar model are determined using finite-temperature field
    theory with auxiliary mass. A new numerical method is developed to
    solve an evolution equation.  The results are discussed in
    comparison with values obtained from the other methods.
\end{abstract}

\end{titlepage}

\newpage
\renewcommand{\thepage}{\arabic{page}}

\section{introduction}\label{intro}

Phase transition is an important phenomenon in particle physics,
cosmology, and condensed matter physics. Quark Gluon Plasma
should be present at the heavy ion collision and will give us a lot of
valuable information on particle physics \cite{QGP}. The investigation
into the chiral phase transition suggest that a number of flavor may
be bounded above \cite{Iwa,Appel}.  In cosmology, the electro-weak
phase transition should be first order for electro-weak
baryogenesis \cite{Coh,Rub} and is investigated
attentively \cite{Car,Arn2,BFH,FH,FHJJM,CFHJJM,GIS,GIKPS,KLRS}. 
Needless to say, a variety of phase transitions are observed and
investigated precisely in condensed matter physics. \\ 

Field theoretical approach is essential in order to investigate these
phase transitions: finite-temperature/chemical-potential field theory
\cite{QGP,Kap}, perturbative and non-perturbative renormalization
group \cite{Wil,MT,DM,AMSST,ABBFTW,BW,Lia}, field theory on lattice
\cite{Mon}, and so on. Temperature can be naturally introduced by
statistical principle using finite-temperature field theory.
Not all the phase transitions, however, can be investigated by it;
the perturbation theory, which is the most powerful method at zero
temperature, often breaks down around the critical temperature because
of many interactions in thermal bath \cite{Sha,Arn3}.
Indeed the perturbation
theory fails, when it is applied to either a second-order or a weakly
first-order phase transition. \\ 

Drummond et.al.\cite{DHL} proposed a new method using an auxiliary mass
in order to avoid this difficulty. We utilized their idea and
developed a new method to calculate the effective potential.  We,
then, investigated the phase transition of the scalar model using the
{\it auxiliary-mass method} and showed it is second order correctly
\cite{A}.  It is a great advance in finite-temperature field theory,
because the phase transition in the scalar model is indicated to be
first order incorrectly by the perturbation theory with daisy
resummation \cite{Arn2,Tak}. We note that the method was able to
reproduce the result with super-daisy approximation \cite{B}. \\ 

Since the equation we must solve in the auxiliary-mass method is a
non-linear partial differential equation for the effective potential,
it can not be solved analytically and must be solved by a numerical
method.  It is, however, difficult to solve partial differential
equations numerically because of the numerical instability \cite{Num}. 
What is worse, the non-linearity of the equation prevents us from
using the methods established in case of a linear equation.  We could
not, therefore, make mesh size arbitrary small; The investigation in
Ref.\cite{A} was not accurate quantitatively.  In the
present paper, we use an improved numerical method given
in the appendix, which do
not suffer from the instability, and get accurate universal
quantities. Unlike the rough values in \cite{A}, they are
beyond the values obtained from Landau approximation.\\ 

The present paper is organized as follows. In the next section we
review the auxiliary-mass method developed in \cite{A}. In section
\ref{res} the effective potential is shown as temperature varies. We,
then, focus on the behaviour of it around the critical temperature and
calculate the universal quantities. These values are compared with
values obtained from the other methods. Summary and discussion is presented
in section \ref{sum}. In appendix we devote to explain the numerical
method we used.\\ 

\section{Review of the auxiliary-mass method}\label{Rev}

We review the method to calculate an effective potential at 
temperature where the perturbation theory is not reliable \cite{A}. We
consider  $\lambda\phi^{4}$ theory which is defined by the
Lagrangian density
\begin{equation}
     {\cal L}_{E}=-\frac{1}{2}
     \left(\frac{\partial \phi}{\partial \tau}\right)^{2}
     -\frac{1}{2}(\mbox{\boldmath $\nabla$} \phi)^{2}
     -\frac{1}{2}m^{2}\phi^{2}
     -\frac{\lambda}{4!}\phi^{4}
     +J\phi + c.t.,
\label{lag}
\end{equation}
where $J$ is an external source function.  If $m^2$ is negative, the
scalar field $\phi$ develops the non-vanishing field expectation value
at $T=0$.  First, the effective potential is calculated with a
positive mass squared $M^2$ which is as large as the temperature
$T^2$. This selection of the mass permit us to use the perturbation
theory without failure, because the loop expansion parameter there is
$\lambda T/M \sim \lambda$ \cite{Arn2,W,Fen}, which is small when the
coupling constant $\lambda$ is small. Using the perturbation theory,
the effective potential is calculated as follows,
\begin{eqnarray}
     V&=&\frac{1}{2}M^{2}\bar{\phi}^{2}
     +\frac{\lambda}{4!}\bar{\phi}^{4}
     +\frac{T}{2\pi^{2}}
     \int^{\infty}_{0}dr r^{2}\log \left[
     1-\exp\left(-\frac{1}{T}\sqrt{r^{2}+M^{2}+
     \frac{\lambda}{2}\bar\phi^{2}}\right)\right].\nonumber \\
    \label{ini}    
\end{eqnarray}
Here, only the one loop thermal correction is left and the quantum
correction is neglected, because it should be negligible when the 
coupling constant $\lambda$ is sufficiently small.\\

We, then, extrapolate the effective potential (\ref{ini}) to the
negative mass squared\footnote{
Hereafter we use unit $\mu=1$. All dimensionful quantities
are measured in the unit.} $m^2=-\mu^2$
 using the following evolution equation,
\begin{eqnarray}
     \frac{\partial V}{\partial m^{2}}&=&
     \frac{1}{2}\bar{\phi}^{2}+\frac{1}{2\pi i}
     \int^{+i\infty +\epsilon}_{-i\infty +\epsilon}dp_{0}
     \int\frac{d^{3}\mbox{\boldmath $p$}}{(2\pi)^{3}}
     \frac{1}{-p_{0}^{2}+\mbox{\boldmath $p$}^{2}+m^{2}
     +\frac{\lambda}{2}
     \bar{\phi}^{2}+\Pi}\frac{1}{e^{\beta p_{0}}-1}\nonumber \\
\label{evo1}
\end{eqnarray}
where $\bar\phi$ is an expectation value of the field and
$\Pi=\Pi(\mbox{\boldmath $p$}^{2}, -p_{0}^{2},\bar{\phi},m^{2},\tau)$
is a full self energy. The thermal correction is left and the quantum
correction is neglected here, too. Of course, $\Pi$ can not be
calculated exactly; we need an appropriate approximation in order to
calculate the effective potential from (\ref{ini}). Because the effective
potential is a generating function of n-point functions with zero
external momentum, neglect of momentum dependence in $\Pi$ allows us
to replace as follows,
\begin{equation}
     m^{2}+\frac{\lambda}{2}\bar{\phi}^{2}+
     \Pi(0,0,\bar{\phi},m^{2},\tau)\rightarrow
     \frac{\partial^{2}V}{\partial\bar\phi^{2}}.
\label{okikae}
\end{equation}
The evolution equation (\ref{evo1}) can be converted to partial
differential equation using this replacement as follows,
\begin{equation}
     \frac{\partial V}{\partial m^{2}}=
     \frac{1}{2}\bar{\phi}^{2}+\frac{1}{4\pi^{2}}
     \int^{\infty}_{0}dr r^{2}\frac{1}{\displaystyle \sqrt{r^{2}
     +\frac{\partial^{2}V}{\partial\bar\phi^{2}}}}
     \frac{1}{\displaystyle \exp\left(\frac{1}{T}\sqrt{r^{2}
     +\frac{\partial^{2}V}{\partial\bar\phi^{2}}}\right)-1}.
\label{evo2}
\end{equation}

The effective potential can be calculated by solving the partial
differential equation (\ref{evo2}) with the initial condition
(\ref{ini}). The effective potential has an imaginary part below the
critical temperature and an analytic continuation is done so that this
imaginary part is negative \cite{A}. Since the evolution equation (\ref{evo2})
is complicated non-linear partial differential equation, it can be
solved only by  numerical methods.

\section{Results}
\label{res}

\begin{figure}
\unitlength=1cm
\begin{picture}(16,6)
\unitlength=1mm
\put(10,35){${\rm Re}\ V$}
\put(120,0){$\phi$}
\centerline{
\epsfxsize=10cm
\epsfbox{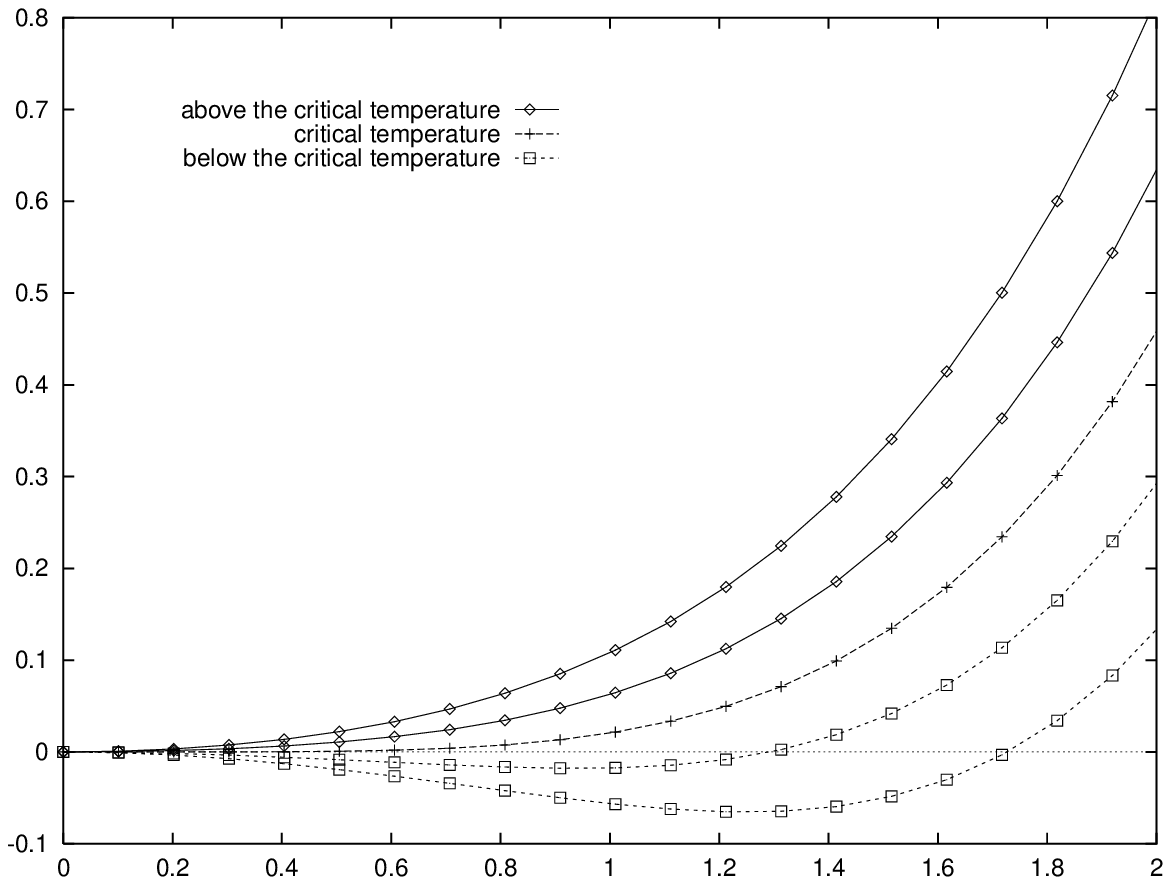} 
} 
\end{picture}
\caption{Real part of the effective potential
($\lambda=1$). The values of the origin are set to zero. A stable
point comes to be zero smoothly as temperature increases.}
\label{potr}
\end{figure}

\begin{figure}
\unitlength=1cm
\begin{picture}(16,6)
\unitlength=1mm
\put(10,35){${\rm Im}\ V$}
\put(120,0){$\phi$}
\centerline{
\epsfxsize=10cm
\epsfbox{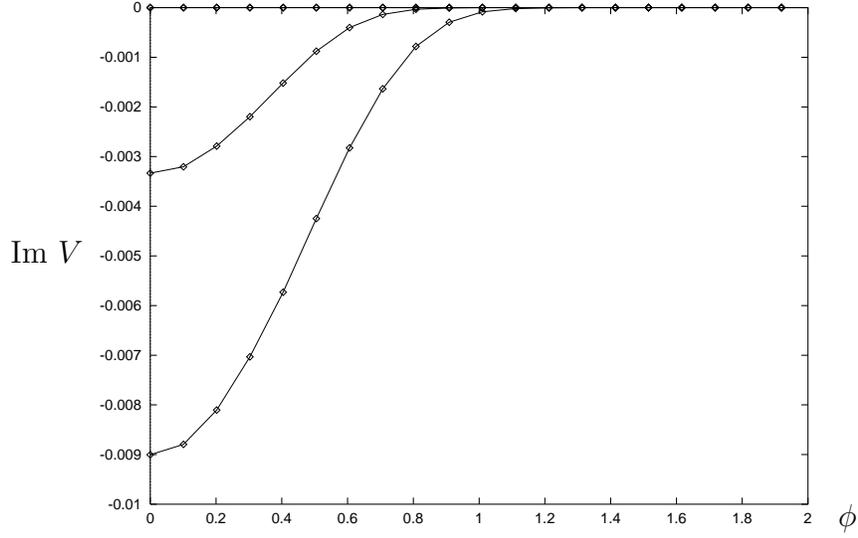} 
} 
\end{picture}
\caption{Imaginary part of the effective potential
($\lambda=1$). The magnitude, which shows the instabity of the state,
increases as a field expectation value decreases. }
\label{poti}
\end{figure}

\begin{figure}
\unitlength=1cm
\begin{picture}(16,6)
\unitlength=1mm
\put(15,35){$T_c$}
\put(120,0){$\lambda$}
  \unitlength=1mm
\centerline{
\epsfxsize=10cm
\epsfbox{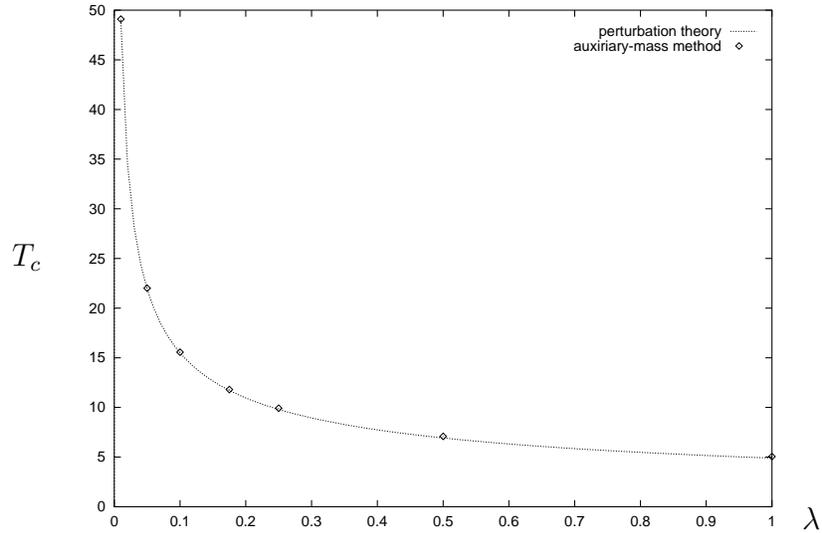} 
} 
\end{picture}
\caption{Phase diagram of $\lambda\phi^4$ theory. Second-order phase
transition is observed on the boundary. The dots represent values
calculated using auxiliary-mass method. The dotted line represents
the leading result of perturbation theory\cite{Dol}.}
\label{critem}
\end{figure}

We calculate the effective potential numerically using the method in
the Appendix. The real part of the effective potential as temperature
varies is shown in fig.\ref{potr}. A stable field expectation value
$\bar\phi_c$, where the effective potential has its minimum, comes to be
zero smoothly as temperature increases. This indicates that
second-order phase transition takes place in this model correctly
\cite{ZJ,Pes}. The imaginary part of the effective potential below the
critical temperature is shown in fig.\ref{poti}. One can observe a
magnitude of it increases as a field expectation value decreases; this
illustrate that a state with smaller field expectation value is less
stable below the critical temperature. The critical temperature as a
function of the coupling constant $\lambda$ is shown in
fig.\ref{critem}. This shows a similar behaviour to the leading result
obtained in Ref.\cite{Dol}, but has a slight difference ($\sim 2 \%$).  In the
remaining of this section we determine some critical exponents:
$\beta , \delta ,\gamma^{+/-}$ ,and $\alpha$. The amplitude ratio
$\chi_+/\chi_-$ is also determined. The results are summarized
in table.\ref{com}.\\ 

\begin{figure}
\unitlength=1cm
\begin{picture}(16,6)
\unitlength=1mm
\put(15,35){$\phi_c$}
\put(120,0){$\tau$}
  \unitlength=1mm
\centerline{
\epsfxsize=10cm
\epsfbox{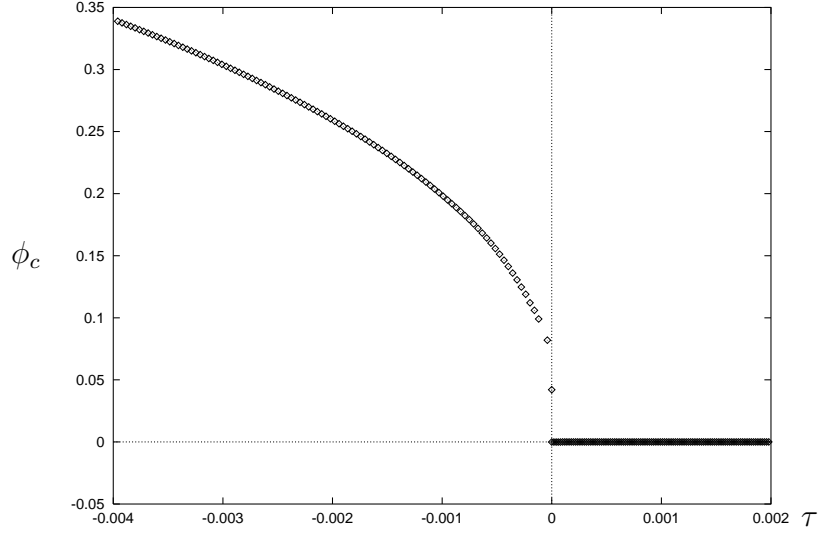} 
} 
\end{picture}
\caption{Stable field expectation value as a function of temperature
($\lambda=1$). It decreases monotonically and vanishes smoothly
as temperature increases.  }
\label{xmin1}
\end{figure}

\begin{figure}
\unitlength=1cm
\begin{picture}(16,6)
\unitlength=1mm
\put(8,35){$\log(\bar\phi)$}
\put(120,0){$\log(-\tau)$}
  \unitlength=1mm
\centerline{
\epsfxsize=10cm
\epsfbox{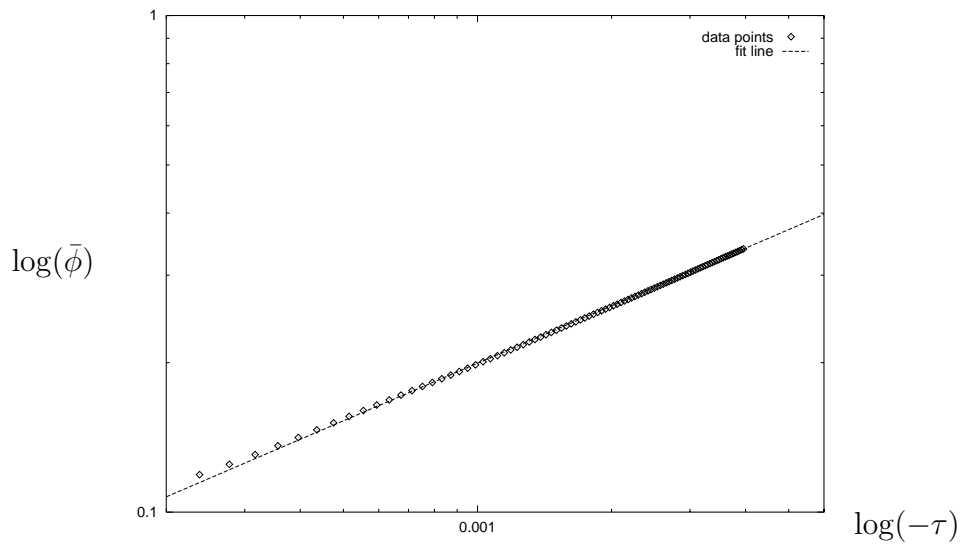} 
} 
\end{picture}
\caption{Plot of $\log(\bar\phi)-\log(-\tau)$ ($\lambda=1$). The
data points are fit to linear function. Using its gradient, $\beta$ is 
determined. }
\label{logxmin}
\end{figure}

First, we observe the stable point $\bar\phi_c$ carefully. 
Figure.\ref{xmin1} shows $\bar\phi_c$ as a function of temperature. It
decreases monotonically and vanish smoothly as temperature increases. 
We, then, focus on its behaviour near the critical temperature
$T_c$ and determine $\beta$, which relate a magnetization to
temperature near $T_c$. This is
defined as follows,
\begin{eqnarray}
     \phi_c\propto (-\tau)^{\beta}\ \ \ (\tau\sim 0, T<T_c)
    \label{beta}
\end{eqnarray}
where $\tau =(T-T_{c})/T_c$. We plot $\log(\bar\phi_c)$
against $\log(-\tau)$ in fig.\ref{logxmin}; we fit the data to linear
function and draw it in fig.\ref{logxmin}. We determine $\beta$ from
the gradient of it. We find $\beta = 0.385$.\\ 

\begin{figure}
\unitlength=1cm
\begin{picture}(16,6)
\unitlength=1mm
\put(15,35){$V$}
\put(120,0){$\phi$}
  \unitlength=1mm
\centerline{
\epsfxsize=10cm
\epsfbox{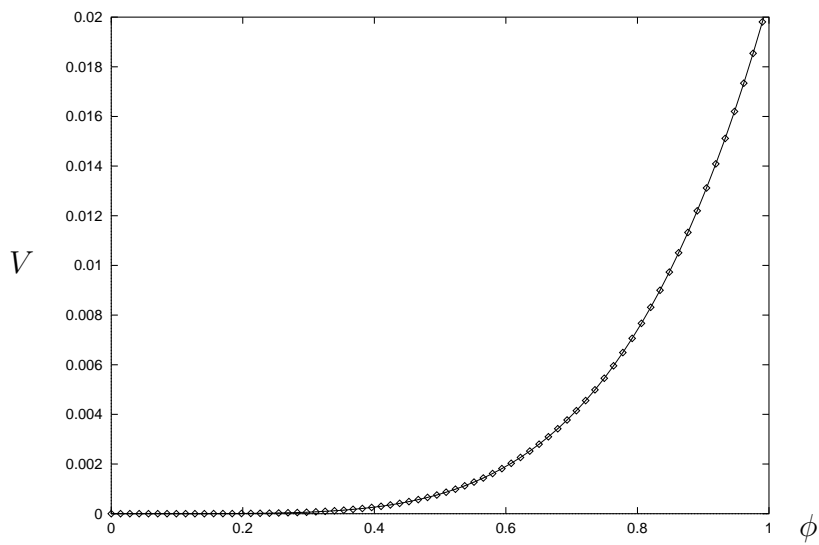} 
} 
\end{picture}
\caption{Effective potential at the critical temperature
($\lambda=1$). }
\label{cripot}
\end{figure}

\begin{figure}
\unitlength=1cm
\begin{picture}(16,6)
\unitlength=1mm
\put(7,35){$\log(V)$}
\put(120,0){$\log(\bar\phi)$}
  \unitlength=1mm
\centerline{
\epsfxsize=10cm
\epsfbox{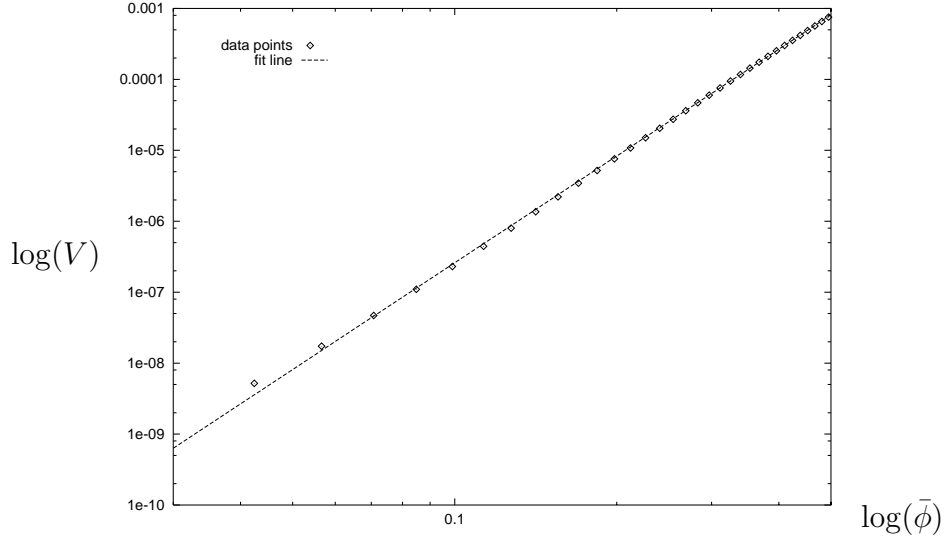} 
}\ 
\end{picture}
\caption{Plot of $\log(V)-\log(\bar\phi)$ ($\lambda=1$). The
data points are fit to linear function. Using its gradient, $\delta$ is
determined. }
\label{logV}
\end{figure}

Next, we determine the exponent $\delta$ which is defined as follows,
\begin{eqnarray}
    \label{delta}
    \bar\phi\propto J^{1/\delta}=(\frac{\partial
    V}{\partial\bar\phi})^{1/\delta} \hspace{1cm}(T=T_c).
\end{eqnarray}
One can derive the following relation from this,
\begin{eqnarray}
    \label{delta2}
    V\propto\bar\phi^{\delta+1}\hspace{1cm}(T=T_c).
\end{eqnarray}
We show the effective potential at $T_c$ in fig.\ref{cripot}. We plot
$\log(V)$ against $\log(\bar\phi)$ in fig.\ref{logV}; we fit
the data to linear function and draw it in fig.\ref{logV}. We
determine $\delta$ from the gradient of it. The result is
$\delta = 4.0$

\begin{figure}
\unitlength=1cm
\begin{picture}(16,6)
\unitlength=1mm
\put(10,35){$\displaystyle
\frac{\partial^2V}{\partial\bar\phi^2}$}
\put(120,0){$\tau$}
  \unitlength=1mm
\centerline{
\epsfxsize=10cm
\epsfbox{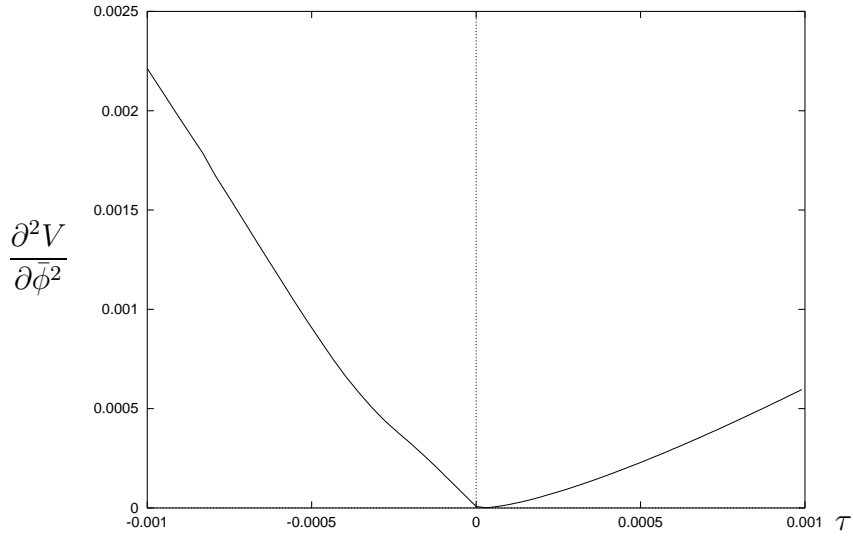} 
} 
\end{picture}
\caption{Curvature at minimum point $\frac{\partial^2
V}{\partial\bar\phi^2}$ as temperature varies ($\lambda=1$).}
\label{cur}
\end{figure}

\begin{figure}
\unitlength=1cm
\begin{picture}(16,6)
\unitlength=1mm
\put(3,35){$\displaystyle \log(\frac{\partial^2 V}{\partial\phi^2})$}
\put(120,0){$\log(|\tau|)$}
  \unitlength=1mm
\centerline{
\epsfxsize=10cm
\epsfbox{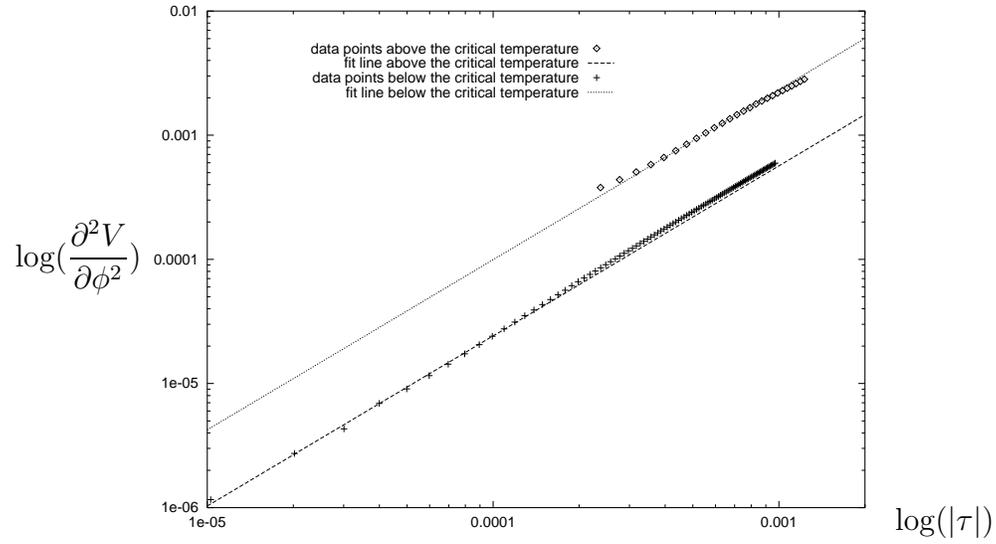} 
} 
\end{picture}
\caption{Plots of $\log(\frac{\partial^2 V}{\partial
\phi^2})-\log(|\tau|)$ ($\lambda=1$). The data points are fit to linear
functions. Using their gradients, $\gamma^{+/-}$ and $\chi_+/\chi_-$ are
determined. }
\label{logcur}
\end{figure}

Then, we determine $\gamma^{+/-}$ and $\chi_+/\chi_-$. They are defined as
follows through the susceptibility,
\begin{eqnarray}
    \label{chi}
    \chi
    &\equiv&
    \left.
    \frac{\partial\bar\phi}{\partial J}
    \right|_{J=0}
    \sim\chi_+\tau^{-\gamma^+}
    \hspace{0.5cm} (\tau\sim 0,T>T_c),\\
    \chi
    &\equiv&
    \left.
    \frac{\partial\bar\phi}{\partial J}
    \right|_{J=0}
    \sim\chi_-\tau^{-\gamma^-}
    \hspace{0.5cm} (\tau\sim 0,T<T_c).
\end{eqnarray}
To calculate $\chi$, we relate $\chi$ to the curvature using the
following identity derived from the definition of the effective
potential,
\begin{eqnarray}
    \label{id}
    \left.
    \frac{\partial\bar\phi}{\partial J}
    \right|_{J=0}
    =
    \left.
    \left(
    \frac{\partial^2 V}{\partial \phi^2}
    \right)^{-1}
    \right|_{\bar\phi=\phi_c}.    
\end{eqnarray}
We show $\left.  \left( \frac{\partial^2 V}{\partial \phi^2}
\right)^{-1} \right|_{\bar\phi=\phi_c}$ as a function of temperature in
fig.\ref{cur}. We also plot $\log(\frac{\partial^2 V}{\partial
\phi^2})$ against $\log(|\tau|)$ in fig.\ref{logcur}; we fit the data
to linear functions and draw them in fig.\ref{logcur}. We determine
$\gamma^{+/-}$ from the gradient of it and
$\chi_+/\chi_-$ from the intercepts.
We find $\gamma \equiv \gamma^+ = \gamma^- =1.37,\ \chi_+/\chi_-=3.4$\\

\begin{figure}
\unitlength=1cm
\begin{picture}(16,9)
\unitlength=1mm
\put(-10,35){$\displaystyle-\frac{\partial^2 V}{\partial T^2}$}
\put(140,0){$\tau$}
  \unitlength=1mm
\centerline{
\epsfxsize=13cm
\epsfbox{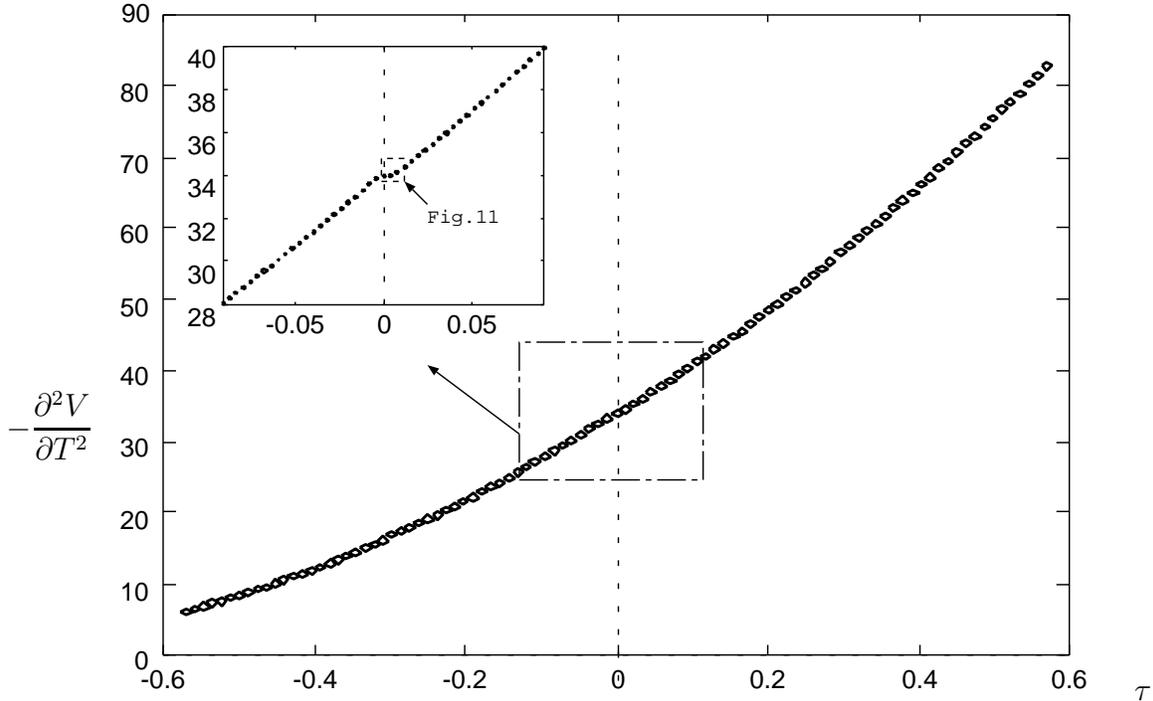} 
} 
\end{picture}
\caption{Second derivative of the effective
potential with respect to temperature\ ($\lambda=1$).
}
\label{cvall}
\end{figure}

\begin{figure}
\unitlength=1cm
\begin{picture}(16,6)
\unitlength=1mm
\put(5,35){$\displaystyle -\frac{\partial^2 V}{\partial\tau^2}$}
\put(120,0){$\tau$}
  \unitlength=1mm
\centerline{
\epsfxsize=10cm
\epsfbox{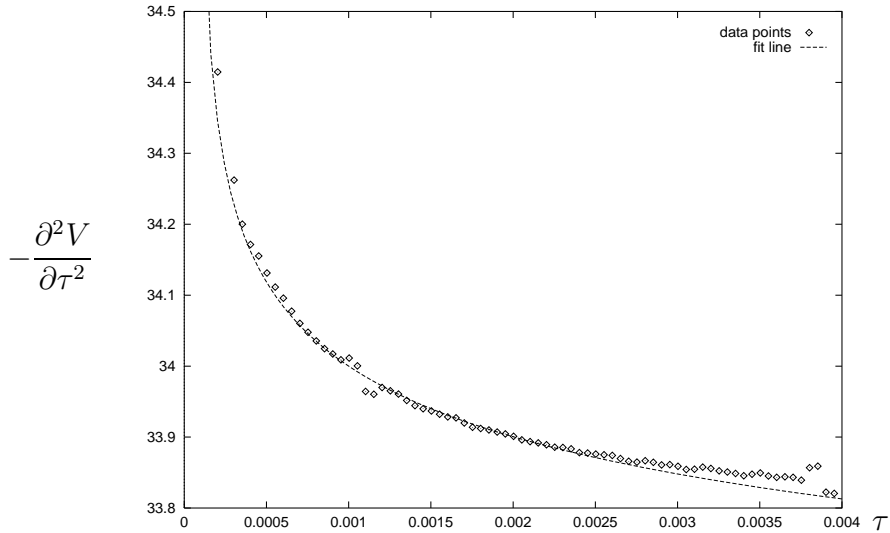} 
} 
\end{picture}
\caption{
Specific heat $C$ as a function of $\tau$ around the critical
temperature. One can observe that it blows up around the critical
temperature. One of the critical exponent $\alpha$ is determined from
this. }
\label{cvcri}
\end{figure}

Finally, we pay attention to the second derivative of the effective
potential with respect to temperature, which is proportional to the specific
heat $C$. 
The exponent $\alpha$ is defined as follows\footnote{
Though the amplitude ratio of the specific heat can also be defined,
it is no determined because of the numerical reason.
},
\begin{eqnarray}
    \label{alpha}
    C\propto\frac{\partial^2 V}{\partial \tau^2}
    \propto\tau^{-\alpha}  \ \ \ (\tau\sim 0).
\end{eqnarray}
This derivative is shown in fig.\ref{cvall} as a function of
temperature. We focus on its behaviour around $T_c$ in fig.\ref{cvcri} 
and observe that it blows up there. One of the critical
exponent $\alpha$ is determined using this. The result is $\alpha=0.12$.

The results are summarized in table.\ref{com} and compared with
results obtained by various methods. Discussion is
presented in the next section. \\

\begin{table}[ht]
{\small
\begin{center}
\begin{tabular}{|c|c|c|c|c|c|c|c|c|c|c|}
\hline
\multicolumn{4}{|c|}{}
&$\gamma$&$\nu$&$\beta$&$\alpha$&$\delta$&$\eta$&$\chi_+/\chi_-$ \\
\hline
 & \multicolumn{3}{|c|}{auxiliary-mass method} &1.37&&0.385&0.12&4.0&&3.4\\
\cline{2-11} 
F-T&perturbation & \multicolumn{2}{|c|}{1-loop} &*&*&*&*&*&*&* \\
\cline{3-11} &theory& \multicolumn{2}{|c|}{2-loop}
&1.0&0.5&0.5&0.0&3.0&0.0&2.0\\
\hline
 &perturbation & \multicolumn{2}{|c|}{fixed dim.} &1.24&0.630&0.325&0.11&4.82&0.317&4.82 \\
\cline{3-11} 
&theory\cite{Gui}& \multicolumn{2}{|c|}{$\epsilon$-exp. }
&1.24&0.631&0.327&0.11& 4.79
&0.349  &4.70\\
\cline{2-11} 
R-G&non& \multicolumn{2}{|c|}{sharp cut off \cite{AMSST} } &(1.38)&0.690&(0.345)&(-0.07)&(5.0)&0.0&* \\
\cline{3-11} &-perturbative &smooth &$\partial^0$&(1.32)&0.660&(0.33)&(0.02)  &(5.0)&0.00&* \\
\cline{4-11}  &   &cut off \cite{MT}&$\partial^2$  
&(1.20)&0.618&(0.327)&(0.146)  &(4.67)&0.054&* \\
\hline
\multicolumn{4}{|c|}{lattice Monte Carlo \cite{HAJ}} 
&1.24&0.629&0.324  & 0.113 &4.83  &0.027 & \\
\hline
\multicolumn{2}{|c|}{} & \multicolumn{2}{|c|}{binary fluids}
&1.236&0.625&0.325&0.112& &  &4.3 \\
\cline{3-11} 
\multicolumn{2}{|c|}{experiment \cite{Gui}} & \multicolumn{2}{|c|}{liquid-vapor} &1.24  & 0.625 &0.316  &0.107  &  &  & 5.0\\
\cline{3-11}\multicolumn{2}{|c|}{} & \multicolumn{2}{|c|}{antiferromagnets} &1.25  &0.64  &0.328  &0.112  &  &  & 4.9\\
\hline
\multicolumn{4}{|c|}{Landau approximation} &1.0&0.5&0.5&0.0&3.0&0.0&2.0\\
\hline
\end{tabular}\\
\end{center}
}
\caption{Critical exponents and  critical amplitude obtained
from various methods. Since first-order phase transition is
indicated, the critical exponents can not be determined using
finite-temperature field theory (F-T) within one-loop order. We note
that there are many non-perterbative methods based on the
renormalization group (R-G) idea which we do not refer here. The
central values of them are shown. Values in parenthesis are determined
using scaling relations.}
\label{com}
\end{table}

\section{Summary and discussion}

\label{sum}

The critical exponents and the amplitude ratio were determined using
the auxiliary-mass method developed in ref.\cite{A} by the improved
numerical method in the appendix. The results are summarized in
table.\ref{com}. We found that  $\lambda\phi^4$ theory
shows second-order phase transition as it should be.
Though the critical exponents calculated here do not satisfy
the scaling relations, they satisfy inequalities of
critical exponents. For example, the inequalities given by
Griffiths\cite{Gri},
\begin{eqnarray}
    \label{ineq}
    \gamma^- &\ge& \beta(\delta-1),\\
    \gamma^+(\delta+1) &\ge& (2-\alpha)(\delta-1)
\end{eqnarray}
are satisfied. In the following we compare our result
with other's.

First, the results are compared with the values obtained by the
perturbative finite-temperature field theory with daisy resummation. 
Since first-order phase transition is indicated at one-loop
order \cite{Arn2,Tak},
the critical exponents can not be determined by the
perturbation theory. At two-loop order, second-order phase transition
is observed and the critical exponents are same as those by the Landau 
approximation \footnote{ We used the two-loop order
effective potential calculated in \cite{Arn2}. We determined the
critical exponents from this both numerically and analytically. }. 
auxiliaryIn comparison with these values, the results obtained in the present paper
are considerably good. \\ 

Second, they are compared with the values obtained by renormalization 
group and by lattice simulation, which agree with greatly. In comparison with
these accurate values, our results are not very  good. These errors 
are probably caused by the replacement (\ref{okikae}). Since this
replacement is based on the neglect of momentum dependence in $\Pi$,
we have to take into account the momentum dependence in order to
improve our results \cite{E}. \\

As mentioned in Sec.\ref{intro}, the finite-temperature field theory
is optimum theory in order to investigate phase transitions; it is
based on statistical principle and can deal with both first-order and
second-order phase transition. The perturbation theory, however, often
breaks down and it prevent us from using the finite-temperature field
theory. The -mass
method enables the finite-temperature field theory to be
used in various situations. \\

We finally express our thanks to T. Inagaki for valuable discussions
and communications. J.S is supported by JSPS Research Fellowships.

\appendix
\section{Numerical method}
\label{app}
\begin{figure}
\unitlength=1cm
\begin{picture}(16,6)
\unitlength=1mm
\put(0,0){
\begin{picture}(16,6)
\unitlength=1mm
\centerline{
\epsfxsize=10cm
\epsfbox{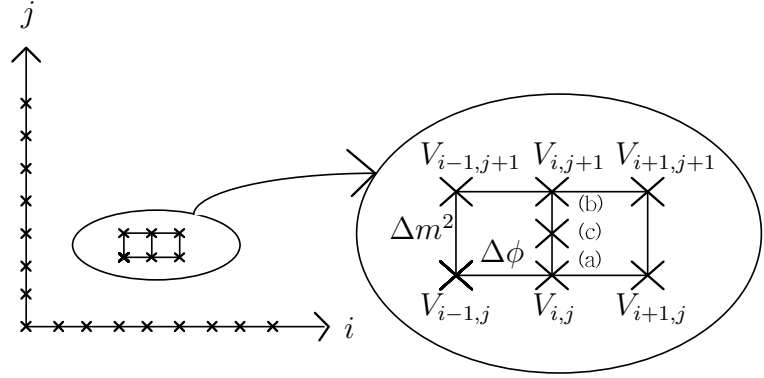} 
} 
\end{picture}
}
\put(22,47){$j$}
\put(65,5){$i$}
\put(75,8){$V_{i-1,j}$}
\put(90,8){$V_{i,j}$}
\put(101,8){$V_{i+1,j}$}
\put(75,28){$V_{i-1,j+1}$}
\put(90,28){$V_{i,j+1}$}
\put(101,28){$V_{i+1,j+1}$}
\put(71,18){$\Delta m^2$}
\put(83,15){$\Delta \phi$}
\end{picture}
\caption{Lattice used to difference (\ref{evo2}). }
\label{lattice}
\end{figure}

The numerical method, which we use to solve (\ref{evo2}), is explained
in this appendix. The partial differential equation (\ref{evo2}) is
written as follows,
\begin{eqnarray}
    \label{def}
    \frac{\partial V}{\partial m^2}=\frac{1}{2}\bar\phi^2 +
    f(\frac{\partial^2 V}{\partial \bar\phi^2}). 
\end{eqnarray}
Here, $f(x)$ is the integral in (\ref{evo2}). First, we make a lattice 
shown in fig.\ref{lattice}. The partial differential equation
(\ref{evo2}) is, then,  differenced as follows \cite{Num},
\begin{eqnarray}
    \label{cra}
    \frac{V_{i,j+1}-V{i,j}}{\Delta m^2}
    &=&
    \frac{1}{2}\phi^2_i+f
    \left(
    \alpha(\frac{V_{i+1,j+1}-2V_{i,j+1}+V_{i-1,j+1}}{(\Delta\phi)^2}
    \right.\nonumber\\ 
    && 
    \left.
    \hspace{2cm}
    +(1-\alpha)(\frac{V_{i+1,j}-2V_{i,j}+V_{i-1,j}}{(\Delta\phi)^2})
    \right).   
\end{eqnarray}
The parameter $\alpha$ decides where the laplacian $\frac{\partial^2
V}{\partial \bar\phi^2}$ is evaluated. If $\alpha=0$ is selected, the
laplacian is evaluated at (a) in fig.\ref{lattice}. The method with
this selection is called the explicit method, which we used in \cite{A}. 
This method is simple, because $V_{x,j+1}$ is determined only by
substituting $V_{x,j}$ into the right hand side. It, however, suffers
from a numerical instability, when smaller mesh $\Delta\phi$ is
chosen \cite{Num}; therefore, we could not make mesh small in
\cite{A}. If $\alpha=1$ is selected, the laplacian is evaluated at (b)
in fig.\ref{lattice}. The method with this selection is called the
implicit method, which does not suffer from the numerical instability
at least if f(x) is a linear function \cite{Num} ----- as far as we
know, when $f(x)$ is not linear function like our case, not many
things are known----- . If $\alpha=1/2$ is selected, the laplacian is
evaluated at (c) in fig.\ref{lattice}. The method with this selection is
called the Crank-Nicholson method, which also does not suffer from the
numerical instability at least if f(x) is a linear function. What is
more, the result converges more rapidly with decreasing $\Delta m^2$ using
this method \cite{Num}. Both the implicit and the Crank-Nicholson
method, however, requires us to solve coupled non-linear equation
(\ref{cra}); this prevents us from using established method in the case
$f(x) \propto x$.
\\ 

We developed two methods in order to overcome this difficulty. First
method is based on the Taylor expansion of $f(x)$. The equation
(\ref{cra}) is rewritten as follows,
\begin{eqnarray}
    \label{cra2}
    \frac{V_{i,j+1}-V_{i,j}}{\Delta m^2}
    &=&
    \frac{1}{2}\phi^2_i+f
    \left(
        \frac{V_{i+1,j}-2V_{i,j}+V_{i-1,j}}{(\Delta\phi)^2}
    \right.\nonumber\\
    && 
    \left.
        +\alpha(\frac{V_{i+1,j+1}-2V_{i,j+1}+V_{i-1,j+1}}{(\Delta\phi)^2}
        -\frac{V_{i+1,j}-2V_{i,j}+V_{i-1,j}}{(\Delta\phi)^2})
    \right).   \nonumber\\ 
\end{eqnarray}
Since the quantity in the parenthesis behind $\alpha$ is the variation of
the laplacian per one step, it is small if $\Delta m^2$ is sufficiently
small. We, then, expand $f(x)$ around
$\frac{V_{i+1,j}-2V_{i,j}+V_{i-1,j}}{(\Delta\phi)^2}$.
\begin{eqnarray}
    \label{cra3}
    \frac{V_{i,j+1}-V_{i,j}}{\Delta m^2}
    &=&
    \frac{1}{2}\phi^2_i+f
    \left(
        \frac{V_{i+1,j}-2V_{i,j}+V_{i-1,j}}{(\Delta\phi)^2}
    \right)\nonumber\\
    && 
    +\alpha
    \left(\frac{V_{i+1,j+1}-2V_{i,j+1}+V_{i-1,j+1}}{(\Delta\phi)^2}
    -\frac{V_{i+1,j}-2V_{i,j}+V_{i-1,j}}{(\Delta\phi)^2}\right)\nonumber\\ 
    &&\ \ \times 
    f^{'}\left(\frac{V_{i+1,j}-2V_{i,j}+V_{i-1,j}}{(\Delta\phi)^2}\right)
    + higher\ order\ terms
\end{eqnarray}
This coupled equation is linear with respect to $V_{x,j+1}$ and can be
solved easily \cite{Num}. \\ 

The second method is based on an iteration. In order to solve equation 
(\ref{cra}), we iterate as follows until a solution is found,
\begin{eqnarray}
    \label{cra4}
    \frac{V_{i,j+1}^{n+1}-V_{i,j}}{\Delta m^2}
    &=&
    \frac{1}{2}\phi^2_i+f
    \left(
    \alpha(\frac{V_{i+1,j+1}^{n+1}-
    2V_{i,j+1}^{n}+V_{i-1,j+1}^{n}}{(\Delta\phi)^2}
    \right.\nonumber\\ 
    && 
    \left.
    \hspace{2cm}
    +(1-\alpha)(\frac{V_{i+1,j}-2V_{i,j}+V_{i-1,j}}{(\Delta\phi)^2})
    \right).   
\end{eqnarray}
Here, n is the number of the iteration.  Note that we can not replace
$V_{i,j+1}^{n}$ with $V_{i,j+1}^{n+1}$ unlike the Gauss-Seidel method,
which is a powerful method if $f(x)$ is a linear function \cite{Num}. 
Next, relaxation method is used in order to improve the convergence
\cite{Num}. Since this procedure is identical with the linear case, we
only write down the iteration equation without an explanation.
\begin{eqnarray}
    \label{cra4}
    \frac{V_{i,j+1}^{n+1}-V_{i,j}}{\Delta m^2}
    &=&
    \omega
    \left(
    \frac{1}{2}\phi^2_i+f
    \left(
    \alpha(\frac{V_{i+1,j+1}^{n+1}-
    2V_{i,j+1}^{n}+V_{i-1,j+1}^{n}}{(\Delta\phi)^2}
    \right.\right.\nonumber\\ 
    && 
    \left.\left.
    \hspace{2cm}
    +(1-\alpha)(\frac{V_{i+1,j}-2V_{i,j}+V_{i-1,j}}{(\Delta\phi)^2})
    \right)\right)   \nonumber\\ 
    &&
    \hspace{0.5cm}+(1-\omega)V_{i,j}^{n}. 
\end{eqnarray}
Here, the relaxation parameter $\omega$ is determined only by
experience. The results by the two methods agree greatly. 
In the present paper, the latter method is used in order to
determine the universal quantities.

\end{document}